\documentclass[conference]{IEEEtran}
\usepackage{amsmath,amssymb,amsfonts}
\usepackage{algorithmic}
\usepackage{graphicx}
\usepackage{textcomp}
\usepackage[table,xcdraw]{xcolor}
\usepackage{rotating}
\usepackage{tikz}
\begin{document}

\title{N=1 Modelling of Lifestyle Impact on Sleep Performance\\
}

\author{\IEEEauthorblockN{Dhruv Upadhyay}
\IEEEauthorblockA{\textit{Donald Bren School of Information and Computer Sciences} \\
\textit{University of California, Irvine}\\
Irvine, USA \\
ddupadhy@uci.edu}
% }
\and
\IEEEauthorblockN{Vaibhav Pandey}
\IEEEauthorblockA{\textit{Donald Bren School of Information and Computer Sciences} \\
\textit{University of California, Irvine}\\
Irvine, USA \\
vaibhap1@uci.edu}
\and
\IEEEauthorblockN{Nitish Nag}
\IEEEauthorblockA{\textit{Donald Bren School of Information and Computer Sciences} \\
\textit{University of California, Irvine}\\
Irvine, USA \\
nagn@uci.edu}
\and
\IEEEauthorblockN{Ramesh Jain}
\IEEEauthorblockA{\textit{Donald Bren School of Information and Computer Sciences} \\
\textit{University of California, Irvine}\\
Irvine, USA \\
jain@ics.uci.edu}
}

% Outline of paper

% So first tell them about event mining, then present data set, and then go into the nitty gritty detail of one explored relationship, tell them specifically what we do and why and then say now look at this nice table that compiles the relevant information for you.

% Create a system block diagram don't spend too much text talking because it gets confusing

% Why are personalized sleep models important??
%Talk about how we do things as well.

%Don't be ambiguous, don't be redundant.
% 

\maketitle

\begin{abstract}
Sleep is critical to leading a healthy lifestyle. Each day, most people go to sleep without any idea about how their night's rest is going to be. For an activity that humans spend around a third of their life doing, there is a surprising amount of mystery around it. Despite current research, creating personalized sleep models in real-world settings has been challenging. Existing literature provides several connections between daily activities and sleep quality. Unfortunately, these insights do not generalize well in many individuals. For these reasons, it is important to create a personalized sleep model. This research proposes a sleep model that can identify causal relationships between daily activities and sleep quality and present the user with specific feedback about how their lifestyle affects their sleep. Our method uses N-of-1 experiments on longitudinal user data and event mining to generate understanding between lifestyle choices (exercise, eating, circadian rhythm) and their impact on sleep quality. Our experimental results identified and quantified relationships while extracting confounding variables through a causal framework. These insights can be used by the user or a personal health navigator to provide guidance in improving sleep.
\end{abstract}

\begin{IEEEkeywords}
sleep quality, event mining, hypothesis verification, N-of-1 experiments, n=1, health, healthy lifestyle, personalized health, precision health.
\end{IEEEkeywords}

% Highlight importance of the problem. Why the approach is novel, what value is this adding to the field of research?

\section{Introduction}
Sleeping well is an essential part of living healthy. The quality of sleep affects the health state of all individuals \cite{Nag2018Cross-modalEstimation} \cite{Nag2020HealthEstimation}. Sleep quality is most often understood as the feeling one has when they wake up in the morning. While it is important to feel refreshed after a night's sleep, the measure of feeling is inherently biased and does not capture much of what sleep research has attributed to sleep quality. Instead of personal perception, other measures of sleep quality require one to take tests that may entail an overnight stay in a sleep lab, known as polysomnography. This could easily bias the results of the sleep quality observed as a person might be uncomfortable in the unfamiliar environment of a sleep laboratory. Although polysomnography is regarded as the gold standard for measuring sleep quality, it is not viable to visit a sleep lab on a nightly basis. If a person would want to know how well they were sleeping, and more importantly, how to improve their sleep quality, they would need to educate themselves on advice from sleep experts. While this is a viable approach, it has a few flaws. For one, the advice given by sleep experts is based on studies done on the general population. These studies have the assumption that the general population shares similar sleeping habits, but it does not account for the cases in which certain people may have different responses to sleep expert advice. This lack of personalization leaves a person using a generalized tool, when they should be using a more sophisticated personalized approach towards lifestyle modification \cite{Nag2017HealthObservations}. For example, a person may know that exercising should help their sleep quality \cite{Kline2014TheImprovement.}, but that would not be able to tell them exactly how much they should exercise to improve their sleep quality significantly, or how varying exercise amount consequently changes its effect on sleep quality. These insights are required if we wish to start controlling our health with more precision using cybernetic principles and health navigation \cite{Nag2017CyberneticHealth} \cite{Nag2019ALife}.

% Include portions about how daily activities can affect sleep quality in the introduction itself.
Sleep is still not fully understood as a science but has many ties to human health. Studies have found sleep's bidirectional connection to immunity \cite{Besedovsky2019TheDisease}. Additional studies have shown the relationship between sleep quality and cardiovascular disease \cite{Cappuccio2017SleepDisease} \cite{Drager2017SleepScience} \cite{Javaheri2017InsomniaDisease}, obesity \cite{Ogilvie2017TheObesity} \cite{St-Onge2017SleepobesityTreatment}, and depression \cite{Steiger2019DepressionSleep}.

It is undeniable that sleep heavily factors into quality of life and has relations to many serious diseases that humans can face. From a computing perspective, these events that impact health and disease risk are known as interface events \cite{Pandey2020ContinuousRetrieval}. It is vital to deepen our understanding of sleep and the factors that affect it by continuously extracting relevant events in our life that modulate sleep. These investigations could lead us to profound discoveries in sleep research and human health. That being said, sleep is an incredibly complex activity that is not easy to understand. The quality and perceived quality can depend on a multitude of factors including mood, stress levels, quality of the mattress, sleeping partners, time of day, etc. These factors each have a more or less significant impact on the quality of sleep depending on the daily activities a person performs. For example, one might expect that after a person exercises for a majority of the day, their physical exhaustion would lead to a good night’s sleep. However, other factors such as heightened stress about the next day might negate the benefits of exercise on sleep quality. Understanding the factors relating to high sleep quality and being able to inform a person about how their lifestyle impacts their sleep quality could be integral to improving quality of life and reducing rates of the aforementioned diseases that are related to poor sleep quality.

We attempt to bridge the gap between the advice of sleep experts and personalized approaches by building a sleep model based on longitudinal data produced by the user. A key factor of our model is to not create general statements as most literature does. Instead, our goal is to show that we can harness the power of event mining to build a personalized model of sleep for a single individual. We record lifestyle choices, such as eating habits, exercise time, previous nights of sleep, and environmental factors via smartphones, wearables, and IoT devices. The model is engineered to provide feedback based on statistically significant causal relationships between daily activities and the coming night of sleep. By introducing this feedback, we can give individuals more understanding about their sleep by enabling them to see the direct effect their choices have on their sleep quality. 

\section{Related Works}
Sleep monitoring and data collection for quality assessments is a growing area of research in the last several decades. This is true in both clinical-grade research applications and consumer-grade applications and products.

\subsection{Sleep Monitoring and Prediction Applications}

The current gold standard for understanding sleep quality is a study known as Polysomnography. This study requires a person to come into a sleep lab or have a sleep expert come to their sleeping location. It is primarily used to diagnose sleep disorders. The test records various metrics such as brain waves, oxygen levels in the blood, heart rate, breathing, and eye and leg movements \cite{PolysomnographyClinic}. This type of study requires a sleep expert and multiple medical sensors. The study's accuracy does come at the cost of needing too many resources and equipment to reliably be performed every night. 

Another popular measurement technique that is used to measure sleep quality is known as Actigraphy. Actigraphy measures sleep quality via a wearable (e.g. a watch). Throughout a sleep event the wearable measures movement with the use of its onboard accelerometer. Actigraphy is a much more accessible form of measuring sleep quality as it only requires the user to remember to wear the device. Its simplicity does come with the cost of accuracy, as it can only infer sleep quality via movement measurements. For the purposes of our study, we used a combination of actigraphy and sound to record the night's sleep events. 

Previous systems that have attempted to perform a similar task have used smartphone mic data in conjunction with machine learning to be able to classify and predict sleep quality \cite{MinTossDetector}. Other studies have attempted to use actigraphy graphs and utilize Deep Learning to predict sleep quality from them as well \cite{Sathyanarayana2016SleepLearning}. There are even studies that attempt to forgo the idea of tracking sleep and use factor graph models based on daily activity to predict sleep quality with 78\% accuracy \cite{Bai2012WillPhone}. While the predictions these models are very powerful, the areas that they lack, and what we want to improve on, deals with giving the user personal feedback to help improve sleep quality. There are also several sleep applications that use audio from the phone mic to record and report sleep quality \cite{SleepClock}\cite{SleepScoreExperts}\cite{SleepTechnologies}. These applications may try to incorporate recommendations for improving sleep quality, but many times the application does not reveal their intuition behind the quality calculation nor does it perform in-depth reviews about how the user can go about improving their sleep as well. These applications are also missing a holistic approach to creating their sleep model as they often only have access to steps taken throughout the day and the last night's sleep to make their recommendations. Our study incorporates multiple lifestyle factors that go beyond the scope of steps taken throughout the day in order to provide more insightful feedback about sleep. 

\subsection{Sleep Quality Measure}

There is currently no agreed-upon true measure of what sleep quality is \cite{Ohayon2016National}. There have been attempts made by sleep researchers to create sleep quality measures regardless. One such measure is the Pittsburgh Sleep Quality Index (PSQI) \cite{SmythMSN2008TheResearch}. Although the PSQI is a highly regarded measure in sleep quality research, it also does involve a user needing to fill out a survey about their sleep events. Responses to these questions can have biased responses. The measure also relies on the fact that a user would be willing to, and would reliably, fill out this survey. For our study, we wanted to use an unbiased measure of sleep; to do so we turned to one of the latest reviews of sleep quality done by sleep experts. 

This review identified the most important factors relating to sleep quality include Sleep Latency (the time it takes to fall asleep), Number of Awakenings Greater than 5 Minutes, Sleep Efficiency (ratio of asleep minutes to minutes in bed), and the Number of Minutes Awake throughout the night \cite{Ohayon2016National}. These four factors flesh out many important factors of a night's sleep that experts believe are essential to attaining high sleep quality. Our model uses these four factors to reason about sleep quality. The review also identified thresholds for sleep quality measures to further define what it means to get a good night's sleep (Table \ref{tab:SleepQualityThresh}).

\begin{table}[]
\begin{center}
\caption{Sleep Quality Measure and Event Thresholds}
\begin{tabular}{|l|l|l|}
\hline
\textbf{Variable}              & \textbf{Classification Ranges}                                                  & \textbf{Event Name}                                           \\ \hline
Sleep Latency                  & \begin{tabular}[c]{@{}l@{}}{[}0, 15{]}\\ (15, 30{]}\\ (30, $\infty$)\end{tabular} & \begin{tabular}[c]{@{}l@{}}Good\\ Average\\ Poor\end{tabular} \\ \hline
Awake Minutes                  & \begin{tabular}[c]{@{}l@{}}{[}0, 20{]}\\ (20, $\infty$)\end{tabular}              & \begin{tabular}[c]{@{}l@{}}Good\\ Poor\end{tabular}           \\ \hline
Awakenings \textgreater 5 mins & \begin{tabular}[c]{@{}l@{}}{[}0, 1{]}\\ (1, $\infty$)\end{tabular}                & \begin{tabular}[c]{@{}l@{}}Good\\ Poor\end{tabular}           \\ \hline
Sleep Efficiency               & \begin{tabular}[c]{@{}l@{}}{[}0.85, 1.00{]}\\ {[}0, 0.85)\end{tabular}          & \begin{tabular}[c]{@{}l@{}}Good\\ Poor\end{tabular}           \\ \hline
\end{tabular}

\label{tab:SleepQualityThresh}
\end{center}
\end{table}

\subsection{Understanding the Effect of Lifestyle Activities on Sleep Quality}

Current literature has made many attempts to identify daily activities that affect a person's sleep quality. Studies have shown that sleep and exercise are related, and greater physical activity levels can lead to better sleep latency \cite{Yang2012ExerciseReview} \cite{Kline2014TheImprovement.}\cite{Kelley2017ExerciseMeta-analyses}. A systematic review has also shown that dietary patterns and the types of food eaten throughout the day lead to better sleep quality and duration \cite{St-Onge2016EffectsQuality}. The environment (temperature and humidity) is also important to our sleep duration and quality \cite{Troynikov2018SleepReview}. From these studies and many more, it is clear that choices made throughout the day have an effect on the quality of the next night's sleep. 

Literature can, and has, told us that certain activities have effects on sleep quality, but it also make these conclusions with controlled environments set in place. The controlled environments are used to eliminate possible confounding variables, which is helpful in showing relationships between daily activities and sleep quality for the general population. While this control is useful, it fails to take into account the complexity of daily life, in which confounding variables cannot be controlled. It becomes much harder to explore the effects that all daily activities would have on sleep quality. Our model uses an event mining approach to help us perform N-of-1 experiments so we can identify causal relationships between daytime activities and sleep quality.

\section{Event Mining}
Our main goal is to model the relationships between sleep and daily activities computationally. To do this modelling, we utilize an analytic technique known as event mining \cite{Pandey2018UbiquitousHealth}. Event Mining is used to find relationships between events present in data. In this work, we attempt to use event mining to find patterns in a user's longitudinal sleep and lifestyle data. Event mining, combined with causal inference principles, allows us to run N-of-1 experiments using a person's data streams. This approach can help us describe relationships between daily activities and create an explainable model of a person's sleep. Event mining will produce rules for us in the form of $Event_i \rightarrow^C Event_o$, where $Event_i$ defines the input event, $Event_o$ is the output event, and C represents the confounding variables and/or the temporal conditions that may affect the relationship between $Event_i$ and $Event_o$. For our experiments, input events will be various daily activities, such as exercise, feeding times, temperature, etc. The output events will be the various sleep quality measures (latency, efficiency, awakenings, and awake minutes). An example of a specific relationship would be $MinutesAwakeBetweenSleepEvents$  $\rightarrow^{Previous Night Awake Minutes} Awake Minutes$. This relationship will further be explored in the Experiments section. 

Hypothesis verification is the process of verifying trends under different scenarios. These patterns are traditionally defined by the literature. Once we have defined our relations between input and output variables, we attempt to identify the confounding variables that affect the relationship significantly. After finding confounding variables, we can perform a technique called contextual matching. This step will find data points that account for similar cases of the confounding variable. From there, we can verify the validity of the relationship and also measure the impact of an input event on the corresponding output event. 

\subsection{Other Techniques}
Another technique we could have used to create a sleep model would be to use Machine Learning to try and fit various models to our data. Machine Learning could indeed create a very powerful model to predict upon the data. An area that could be lacking with this approach is relationship verification. In a complicated model, it becomes difficult to keep track of how input variables relate to each other and how they eventually affect the outcome of the model. Some advantages of event mining are that it allows us to efficiently test and verify relationships between various confounding variables and to directly measure the impacts of the input event on the output event. 

\section{Methods}

\begin{figure*}[htbp]
\centerline{\includegraphics[scale=0.45]{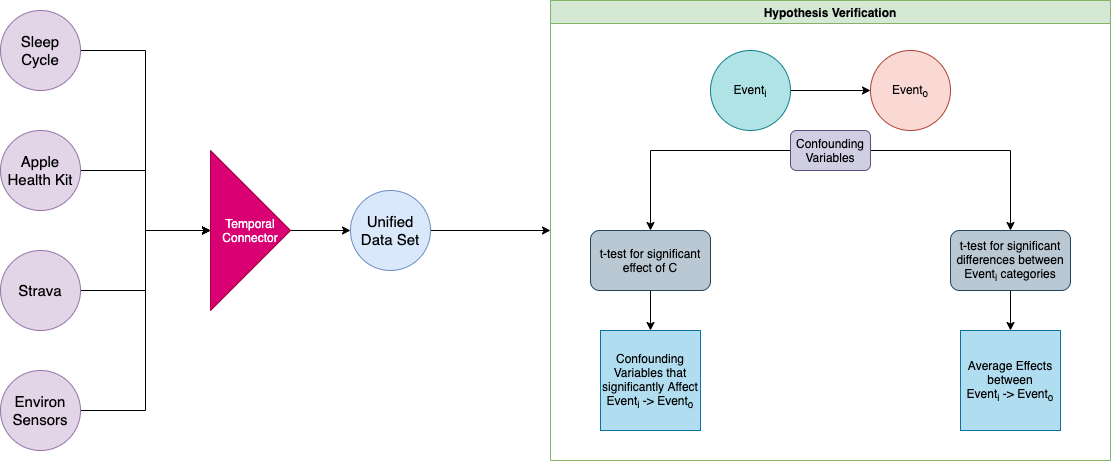}}
\caption{Modeling Pipeline: This diagram describes the pipeline with which we take in the data from all of our sources, combine it into a unified one via temporal relations and then pass it through an event mining process to verify insights and generate the effects that input events have on output events.}
\label{fig:model}
\end{figure*}

This section will explore our data and apply it to create a model that can give a user personalized feedback about their sleep quality. Figure \ref{fig:model} shows the basic outline of the process we will go through to perform our experiments. 

\subsection{Data Set}
The data set we used was made by temporally merging multiple data sources in order to help us better understand the connection between input and output events with regards to sleep quality. 

In order to perform a holistic review of a person's daily activities and their effects on their sleep, we recorded the following pieces of data over the course of multiple years. Daily data were gathered from four main sources: Sleep Cycle, Apple Health Kit, Strava, and Environmental Sensors. The devices used to create these data sets include the user's Garmin Fenix 5 Smart Watch, their smartphone, and an IoT sensor. Sleep Cycle was primarily used to keep track of sleep events. Apple Health Kit was used to help compile sleep quality measures recorded by the Garmin Smart Watch, daily step counts, and floors climbed. The actigraphy measures of the smartwatch were combined with the sound recordings of Sleep Cycle in order to create sleep quality measures. Strava was used to keep track of exercise events. Environmental Sensors were used to keep track of feeding times (phone camera metadata), and temperature and humidity, and the start of sleep events (IoT Sensor). All of these data sources were then temporally matched in order to accurately record lifestyle events that took place throughout the day. 

Additionally, some preprocessing was done on the data to ensure that the data set consisted of consecutive days in terms of sleep recordings. This ensures that we would be able to accurately gather information about the daily activities that would then directly affect the next night's sleep.

In order to make the data easier to work with in terms of an event mining framework, we set up classification ranges for our important data points. The classifications in Table \ref{tab:SleepQualityThresh} and Table \ref{tab:my-table2} are both used to transform our data from a continuous to discrete format.

\begin{table}[]
\begin{center}
\caption{Lifestyle Factors and Event Thresholds}
\begin{tabular}{|c|l|l|}
\hline
\textbf{Variables}                                                             & \multicolumn{1}{c|}{\textbf{Classification Ranges}}                                         & \textbf{Event Name}                                                           \\ \hline
\begin{tabular}[c]{@{}c@{}}Exercise Minutes\\ Per Day\end{tabular}             & \begin{tabular}[c]{@{}l@{}}{[}0{]}\\ (0, 50{]}\\ (50, 150{]}\\ (150, $\infty$)\end{tabular} & \begin{tabular}[c]{@{}l@{}}None\\ Poor\\ Average\\ Good\end{tabular} \\ \hline
\begin{tabular}[c]{@{}c@{}}Exercise Minutes \\ Per Week\end{tabular}           & \begin{tabular}[c]{@{}l@{}}{[}0, 150{]}\\ (150, 300{]}\\ (300, $\infty$)\end{tabular}       & \begin{tabular}[c]{@{}l@{}}Poor\\ Average\\ Good\end{tabular}        \\ \hline
\begin{tabular}[c]{@{}c@{}}Interval Between\\ Eating and Sleeping\end{tabular} & \begin{tabular}[c]{@{}l@{}}{[}0{]}\\ (0, 180{]}\\ (180, $\infty$)\end{tabular}              & \begin{tabular}[c]{@{}l@{}}Missing\\ Poor\\ Good\end{tabular}        \\ \hline
\begin{tabular}[c]{@{}c@{}}Minutes Awake\\ Between Sleep Events\end{tabular}   & \begin{tabular}[c]{@{}l@{}}{[}0, 900{]}\\ (900, 1020{]}\\ (1020, $\infty$)\end{tabular}     & \begin{tabular}[c]{@{}l@{}}Poor\\ Average\\ Good\end{tabular}        \\ \hline
Starting Temperature                                                           & \begin{tabular}[c]{@{}l@{}}{[}0, 60{]}\\ (60, 67{]}\\ (67, $\infty$)\end{tabular}           & \begin{tabular}[c]{@{}l@{}}Cold\\ Comfortable\\ Warm\end{tabular}    \\ \hline
Starting Humidity                                                              & \begin{tabular}[c]{@{}l@{}}{[}0, 30{]}\\ (30, 50{]}\\ (50, 100{]}\end{tabular}              & \begin{tabular}[c]{@{}l@{}}Low\\ Ideal\\ High\end{tabular}           \\ \hline
\end{tabular}
\label{tab:my-table2}
\end{center}
\end{table}

% Go into one of the experiments in good depth, then we can make a large table and just display stuff

\subsection{Experiments}
For the experiments, there are ten input/confounding events that can be used: Previous Night's Sleep Quality Measures (4 categories), Exercise Minutes in the Day, Exercise Minutes Per Week, Interval Between Eating and Sleeping, Minutes Awake Between Sleep Events, Starting Temperature, and Starting Humidity. The possible output events are sleep quality measures (Table \ref{tab:SleepQualityThresh}). Since there are many experiments that we can perform with our possible input, output, and confounding variable tuples, we will go over one such experiment with close detail to explain the process. Afterward we display various figures containing all of our results. After displaying our results, we will go over any significant observations that we were able to find. 

\begin{figure*}[htbp]
\centerline{\includegraphics[width=0.95\linewidth]{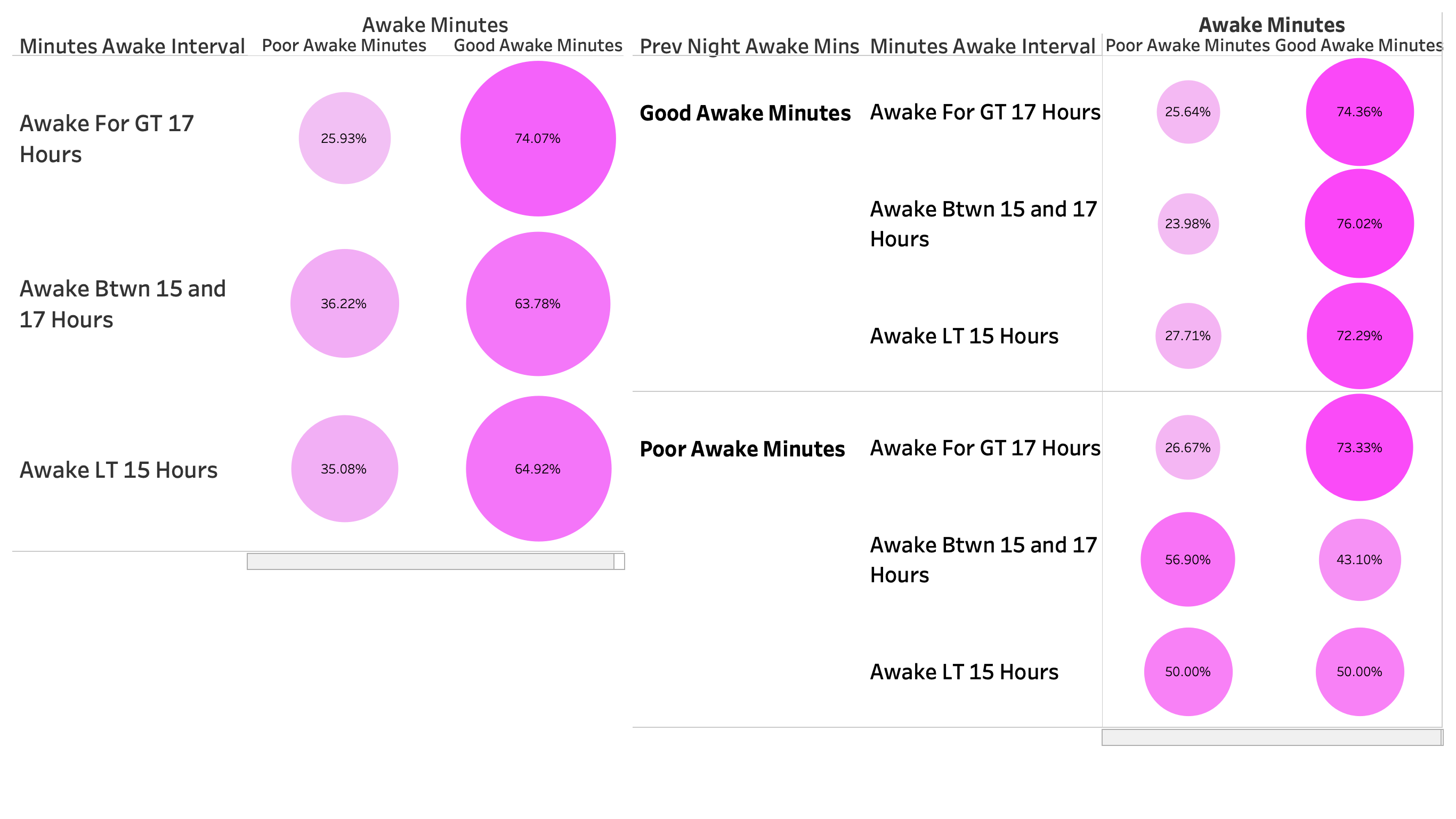}}
\caption{This figure shows the Baseline Distribution of Minutes Between Sleep Events  and Awake Minutes (left) and then shows how the distribution changes when it is conditioned on each of the categories of Prev Night Awake Mins (right).}
\label{fig:baselineVScond}
\end{figure*}

Our experiments have 2 stages. The first stage is designed to find any confounding variables that may affect the relation between and input and output event. The second stage is designed to find the average numerical effect that a certain $Event_i$ will have on $Event_o$, regardless of the confounding variables. Both stages utilize the power of Welch's t-tests to determine statistical significance. For all experiments, a p-value of 0.05 will be used as an indicator of statistical significance. For this section, we will specifically explore analyzing $Event_i$ = Minutes Awake Between Sleep Events, $Event_o$ = Awake Minutes, and $C$ = Prev Night Awakening Minutes. Since the Prev Night Awakening Minutes category has 2 categories in it, we will proceed to condition on each category, and then analyze how it changes the relation between $Event_i$ and $Event_o$. Figure \ref{fig:baselineVScond} shows the baseline distribution of $Event_i$ vs. $Event_o$ in the form of a heat map (left); the conditioned heat maps are shown as well (right).

Based on Figure \ref{fig:baselineVScond}, we can see that when the Previous Night had a poor sleep quality measure for awake minutes. The distribution of the data changes quite drastically when the person was awake between 17 hours and less ([0, 1020] minutes). To verify this, we can perform a t-test between the baseline distribution and the distribution conditioned on the confounding variable. A summary of the t-tests performed can be seen in Table \ref{tab:PrevAwakeCombo}. As suspected, a poor quality measure for last night's Awake Minutes significantly changes the effect that $Event_i$ has on $Event_o$, especially when the individual has been awake for less than 17 hours. Interestingly enough the distributions, and lack of significance from the t-test, seem to suggest that whatever relationship exists between staying awake for greater than 17 and $Event_o$ is strong enough to withstand the influence of having poor sleep last night. An additional insights we can gain from this t-test is that we can confirm that last night's sleep event has a strong influence on the relationship between $Event_i$ and $Event_o$ in two significant cases. This form of analysis gives us the ability to quickly identify any confounding variables that might affect a causal relationship between a pair of input and output events.

The next step of this experiment was to perform a similar t-test against the distributions when conditioning on having a good Previous Night Awake Minutes. A summary of the t-test results can be seen in Table \ref{tab:PrevAwakeCombo}. These t-tests further confirm that last night's sleep quality is an important confounding variable with regards to the relationship between $Event_i$ and $Event_o$. We can see via the t-tests that last night's sleep seems to boost the chance of getting a good night's sleep the next evening. This insight actually holds for most other input-output event pairs as well. A high-level overview of all confounding variables that we found significantly affecting the relationship between various input-output event pairs can be found in the appendix (Figures \ref{fig:LatPart1} - \ref{fig:EffPart2}). This image reduces the complexity of the results by only displaying the p-values of the most significant t-test that changes the distribution between input and output events. For example, in the experiment we ran above, two t-tests were run with respect to the baseline distribution of being awake between 15 to 17 hours and the conditioned ones. In any of the figures mentioned, the p-value displayed would be the more significant p-value between those two tests. In general, the p-value displayed represents the most significant effect that the confounding variable could have on the relationship between $Event_i$ and $Event_o$. Additionally, it is important to note that, in these tables any colored square represents a significant relationship. The larger the square, the more significant the relationship is. Before we talk about other interesting insights that could be gathered from these figures, we will go through the second stage of our experiments using the same input and output events. 

\begin{table}[]
\caption{Summary of t-test results between Baseline Distribution of Minutes Between Sleep Events vs. Minutes Awake and the same distribution conditioned on quality of Awake Minutes the previous night.}
\begin{tabular}{|l|l|l|l|}
\hline
\textbf{Condition}                                                           & \textbf{Distribution} & \textbf{\begin{tabular}[c]{@{}l@{}}Mean Difference \\From Base \\ When Conditioned \\on Previous\\ Night Awake \\Minutes\end{tabular}} & \textbf{\begin{tabular}[c]{@{}l@{}}Significant\\ According\\ p = 0.05?\end{tabular}} \\ \hline
\begin{tabular}[c]{@{}l@{}}Good \\Previous Night\\ Awake Minutes\end{tabular}  &                       &                                                                                                                                  &                                                                                      \\ \hline
                                                                             & (1020, $\infty$)      & +1.45 Minutes                                                                                                                    & No                                                                                   \\ \hline
                                                                             & (900, 1020{]}         & -5.23 Minutes                                                                                                                    & Yes                                                                                  \\ \hline
                                                                             & {[}0, 900{]}          & -4.30 Minutes                                                                                                                    & Yes                                                                                  \\ \hline
\begin{tabular}[c]{@{}l@{}}Poor \\Previous Night \\ Awake Minutes\end{tabular} &                       &                                                                                                                                  &                                                                                      \\ \hline
                                                                             & (1020, $\infty$)      & -3.78 Minutes                                                                                                                    & No                                                                                   \\ \hline
                                                                             & (900, 1020{]}         & +8.84 Minutes                                                                                                                    & Yes                                                                                  \\ \hline
                                                                             & {[}0, 900{]}          & +8.71 Minutes                                                                                                                           & Yes                                                                                  \\ \hline
\end{tabular}
\label{tab:PrevAwakeCombo}
\end{table}

\begin{figure*}[htbp]
\centerline{\includegraphics[width=0.95\linewidth]{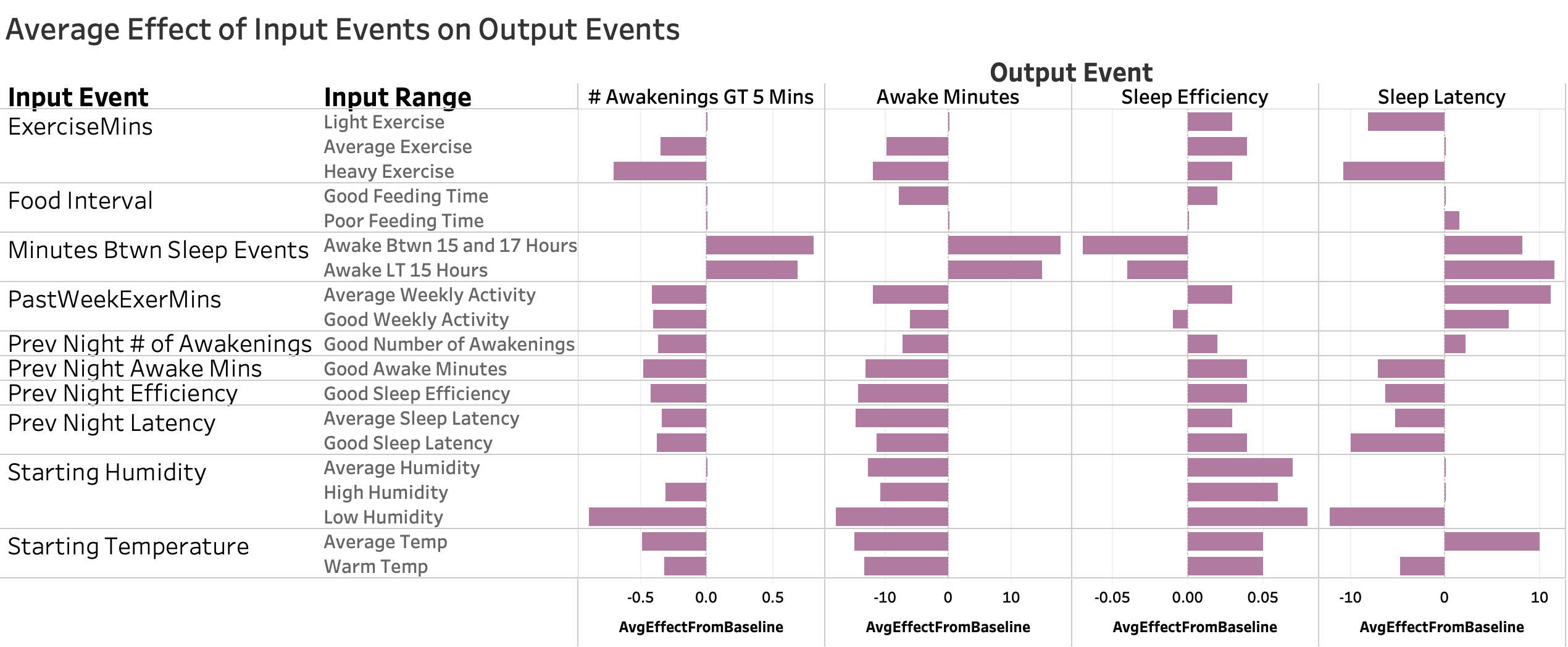}}
\caption{Average Effects that each input event has on the output event when compared to each input event's base category. If a metric is 0 then no significant relations were found.}
\label{fig:AvgEffects}
\end{figure*}

The second stage is focused on presenting the user with feedback about how variations in their daily activities specifically affect their sleep quality on average. To begin, we will once again refer to Figure \ref{fig:baselineVScond}, specifically the bottom right heat map. As can be seen from the figure, the distribution for awakening when sleeping after being awake for greater than 17 hours is drastically different when compared to sleeping after staying awake for less than 17 hours. For this experiment, we treat the first category of $Event_i$ as the baseline and then compare it to the other categories that are present in $Event_i$. If the t-test shows a statistical significance, then we know that under the condition of $C$, there is a significant difference in the distribution between the base category for $Event_i$ and the tested category. For a given $Event_i \rightarrow^C Event_o$ combination we perform these t-tests between the base category and the rest of the categories in $Event_i$ in order to find out how, on average, changing the input of $Event_i$ affects the outcome of $Event_o$. If we similarly carried out this experiment for all other nine confounding events and then averaged over all of the significant mean differences, we would find that, on average, staying awake for 15 to 17 hours when compared to the baseline, staying awake greater than 17 hours, would increase the minutes awake during a sleep event by 18 minutes. Similarly, when we compare the baseline event of staying awake greater than 17 hours to staying awake less than 15 hours, we can see that, on average, the minutes awake during the sleep event will increase by 15 minutes. Since both test conditions show that, compared to the baseline, staying awake less than 17 hours seems to detract from sleep quality we can say that staying awake greater than 17 hours seems to improve sleep quality.

The results of these experiments can be inspected in Figure \ref{fig:AvgEffects}. One interesting observation to note is that if we hold $Event_i$ the same and attempt to analyze its average effect on output events, we can see that an average temperature(60-67 $F^o$) seems to improve every sleep quality measure except for sleep latency. This is a profound observation as it shows that not all quality measures are correlated with each other and that an improvement in one given a certain input event does not necessarily equate to an improvement in all other sleep quality measures. Another example of this phenomenon can be seen with regards to the input event of the count of the number of Awakenings greater than 5 minutes for the previous night. Where compared to the baseline of waking up more than once for greater than 5 minutes, getting a good sleep quality measure seems to improve only 3 out of the 4 sleep quality measures for the next night. Once again, sleep latency deteriorate while all other sleep quality measures get improved.

Another interesting metric that should be noted is that exercise improves sleep latency the most. On average, we can tell that exercising a lot will reduce sleep latency by 10.5 minutes with just a small workout will help reduce sleep latency by an average of 8 minutes. This model can now give evidence-backed feedback to its user about how daily activities affect their sleep quality and even predict how much certain lifestyle changes will transform their sleep quality. 

If we further analyze the Figures present in the appendix, we can also generally see that the last night's sleep is a significant confounding variable to consider in the sleep model. Another strong confounding variable is the amount of time elapsed between sleep events. It seems that longer times spent awake yield good results with regards to sleep quality. With specific regards to sleep efficiency, it is interesting to note that starting humidity and temperatures seem to be very important confounding variables. Additionally, we can see that exercise is not as significant of a confounding variable as the other daily activities, which could indicate that physical exhaustion might not be an important variable to consider when analyzing sleep quality.

\section{Conclusion}
Throughout this paper, we have shown the need for and built up a sleep model that utilizes event mining to provide useful feedback about the relationships between sleep quality. With enough data, this model can be very powerful and give people control over their sleeping habits in a way that has not been previously possible. This model has the power to identify various relationships between lifestyle choices and sleep quality and the possible variables that could strengthen or weaken those relationships.

Using our data and our event mining approach, we were able to identify various factors and how they affect the quality of sleep. For example, one such finding showed that exercising more than 150 minutes on a day that a person has had a good previous night's sleep can, on average, decrease the next night's sleep latency by about 10.5 minutes. Further analysis can show that the starting temperature for the sleep event can significantly strengthen the relationship between exercise and sleep, whereas having poor sleep the night before can detract from the beneficial effects of exercising. While these findings might not generalize well to the population, they do not have to. The power of the model comes from the fact that it is unique to the person. That being said, the model is made such that it can fit another person's data just as easily. Given another person's data set, the model can perform similar analyses and find different insights into how their daily activities affect their sleep quality.

\section{Future Directions}
Although this model incorporates many useful data sources and provides insights about them, it is also a proof of concept. The input events were all based on variables that literature had already identified to be important to sleep quality. While we were able to introduce further levels of complexity to the model to provide more insights than the literature does, there are also many more input events that we would need to include to create a comprehensive sleep model. The biggest missing piece from our current model is an input event that deals with stress and anxiety. Literature has found that anxiety levels do indeed affect sleep quality \cite{Gould2017AssociationSleepiness}. Intuitively, it makes sense that heightened anxiety would lead to difficulty falling asleep and could even cause more awakenings during a sleep event. Additionally, while it is possible to analyze this data and present results, a lot more work would need to be done before a user could plug in various data sources and subsequently get a succinct presentation about how their lifestyle affects their sleep quality. Moving forward, we hope to incorporate even more input events to extend to personal events, emotions, caloric intake, and heart rate. With these inputs, we would hopefully be able to find more insights and even identify certain lifestyle factors that affect sleep quality that has not been considered by the literature yet. That being said, the model is built around a plug and play framework, so incorporation of these and any other future features should be easily doable.

\bibliographystyle{plain}\bibliography{references}

\section{Appendix}
See following pages for Figures 4 through 11.

\begin{sidewaysfigure*}
\centerline{\includegraphics[scale=0.16]{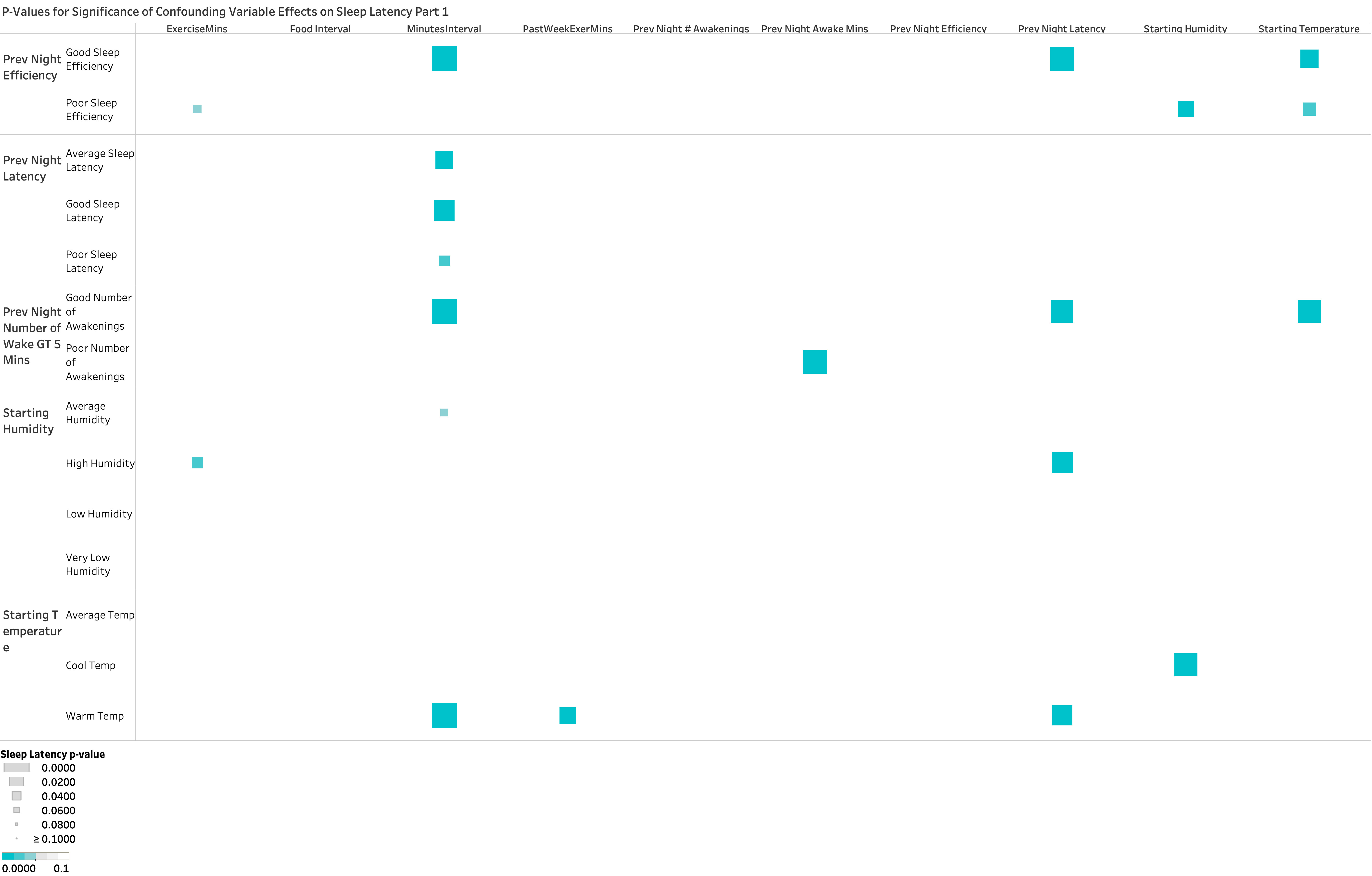}}
\caption{This figure identifies the confounding variables (columns) that affect the relationship between the second half of various input events (rows) and Sleep Latency. The colored squares indicate a significant p-value. The larger the square the more significant the relationship was found.}
\label{fig:LatPart1}
\end{sidewaysfigure*}

\begin{sidewaysfigure*}
\centerline{\includegraphics[scale=0.175]{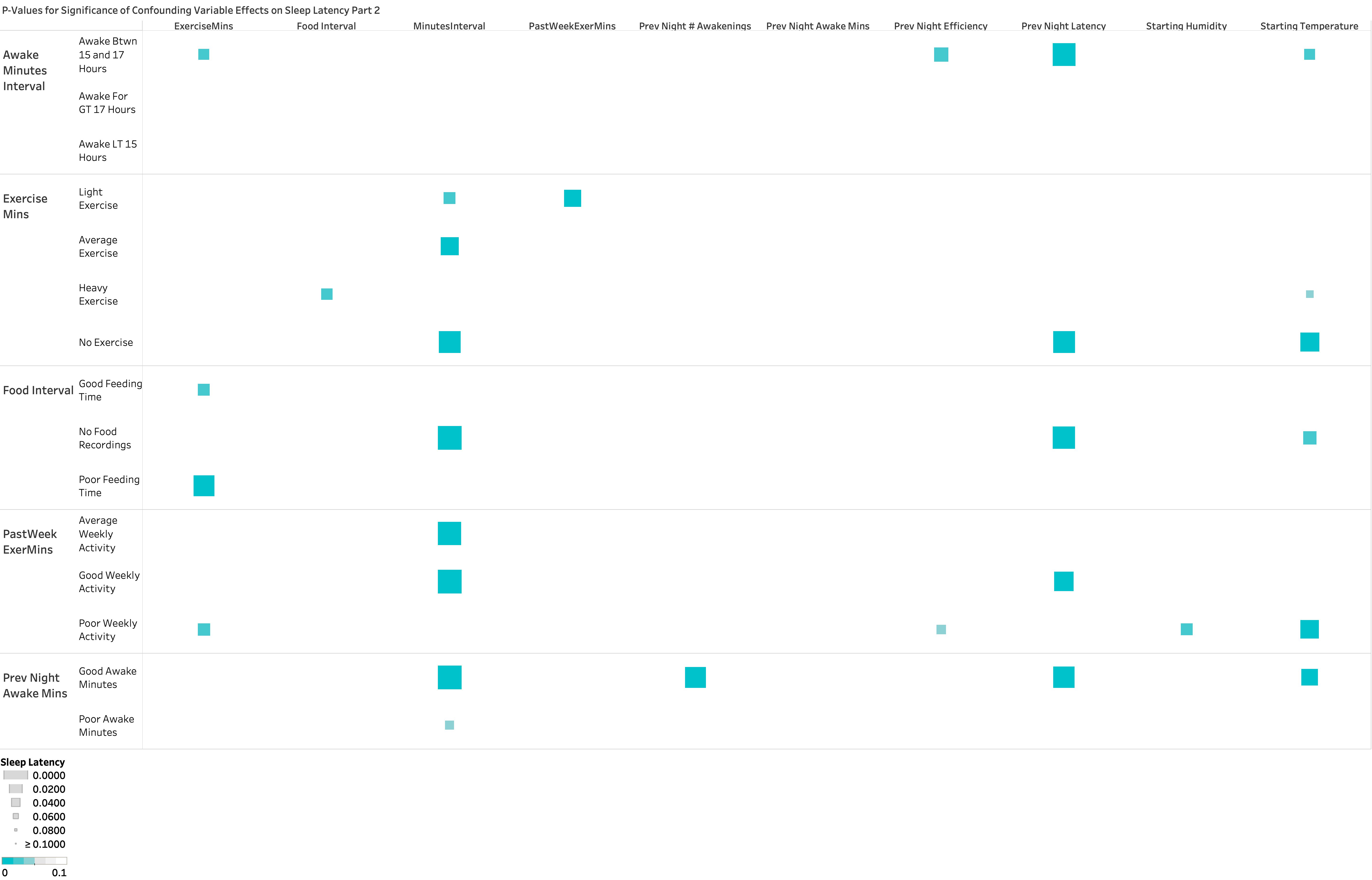}}
\caption{This figure identifies the confounding variables (columns) that affect the relationship between the second half of various input events (rows) and Sleep Latency. The colored squares indicate a significant p-value. The larger the square the more significant the relationship was found.}
\label{fig:LatPart2}
\end{sidewaysfigure*}

\begin{sidewaysfigure*}
\centerline{\includegraphics[scale=0.175]{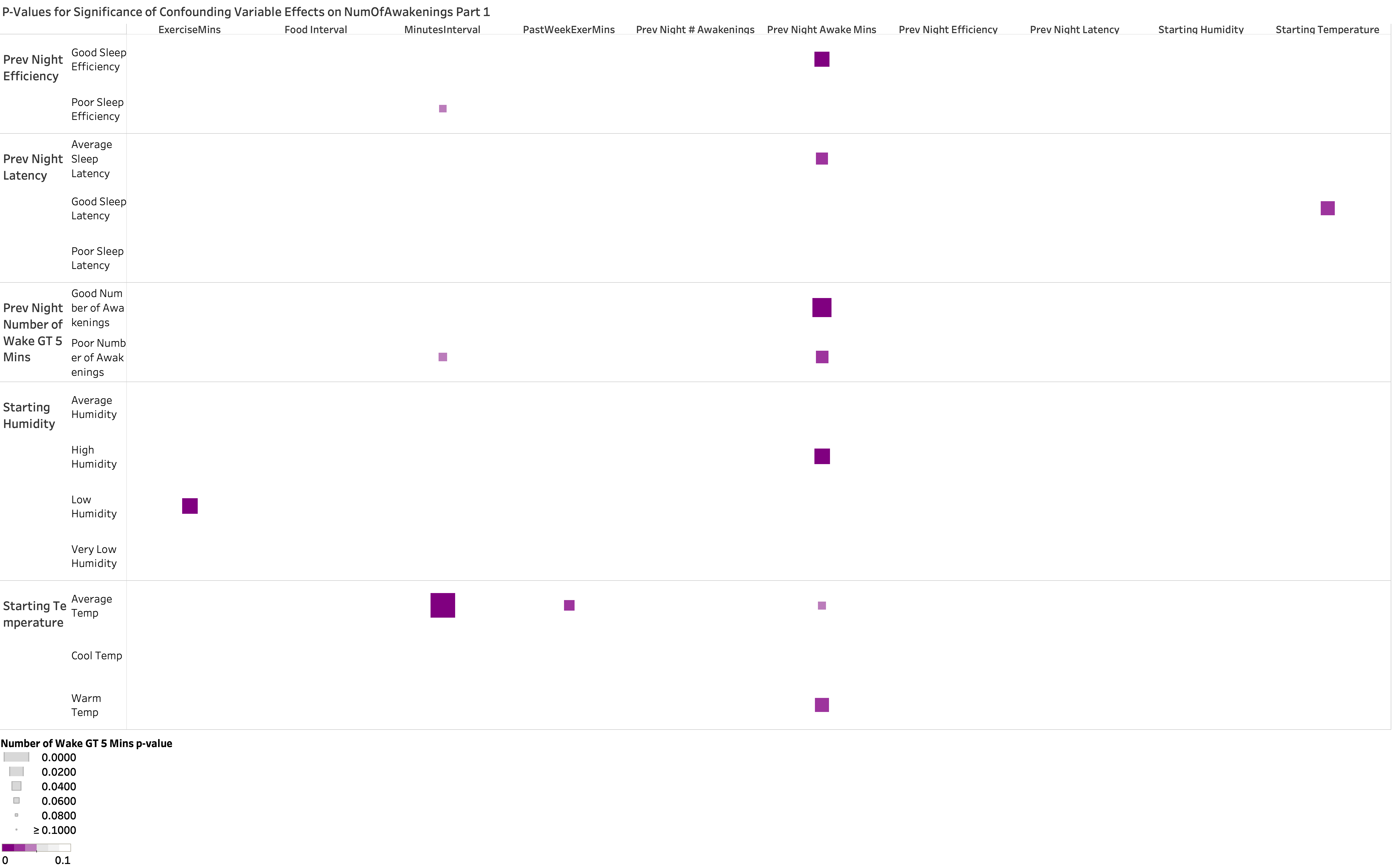}}
\caption{This figure identifies the confounding variables (columns) that affect the relationship between the first half of various input events (rows) and Number of Awakenings GT 5 Mins. The colored squares indicate a significant p-value. The larger the square the more significant the relationship was found.}
\label{fig:NumPart1}
\end{sidewaysfigure*}

\begin{sidewaysfigure*}
\centerline{\includegraphics[scale=0.175]{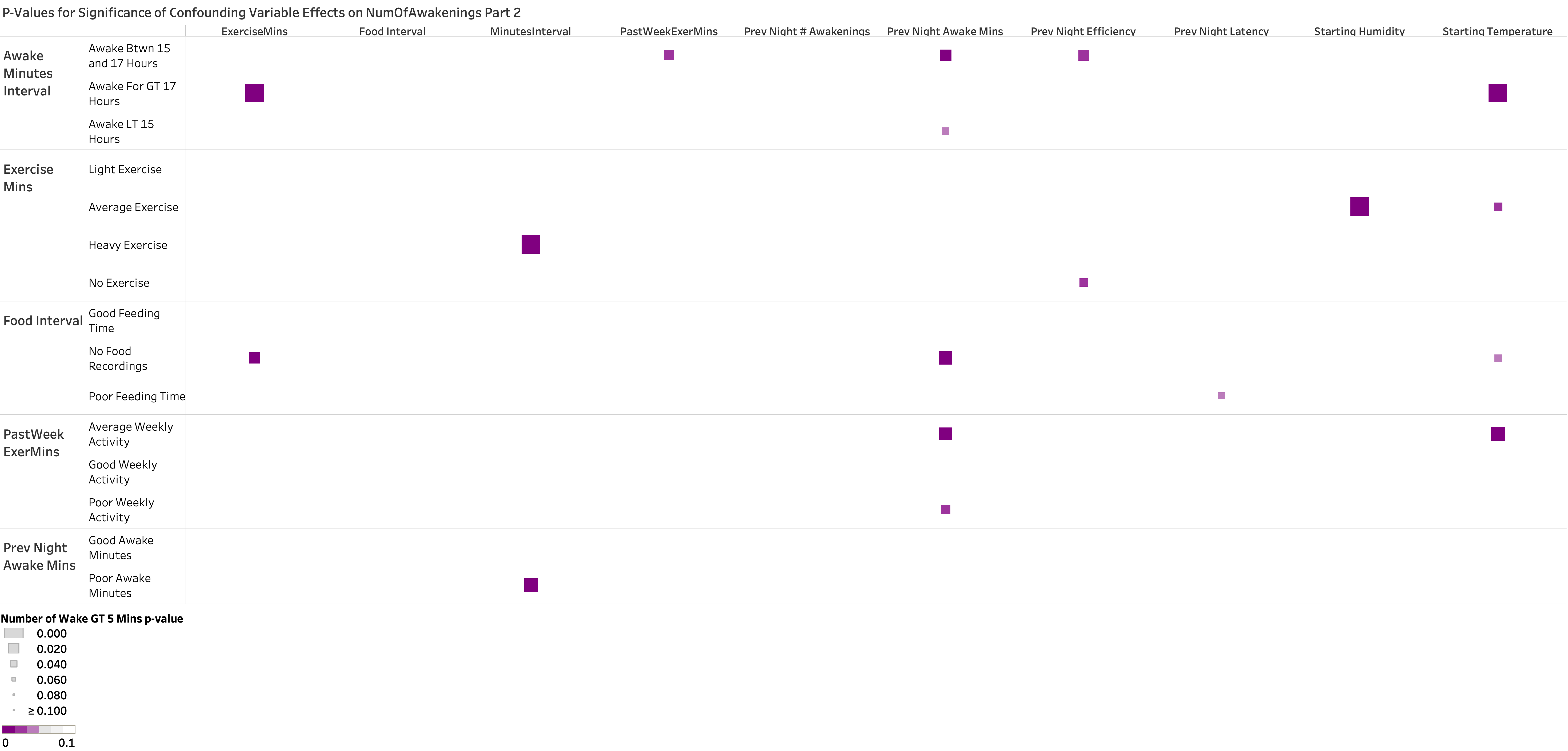}}
\caption{This figure identifies the confounding variables (columns) that affect the relationship between the second half of various input events (rows) and Number of Awakenings GT 5 Mins. The colored squares indicate a significant p-value. The larger the square the more significant the relationship was found.}
\label{fig:NumPart2}
\end{sidewaysfigure*}

\begin{sidewaysfigure*}
\centerline{\includegraphics[scale=0.175]{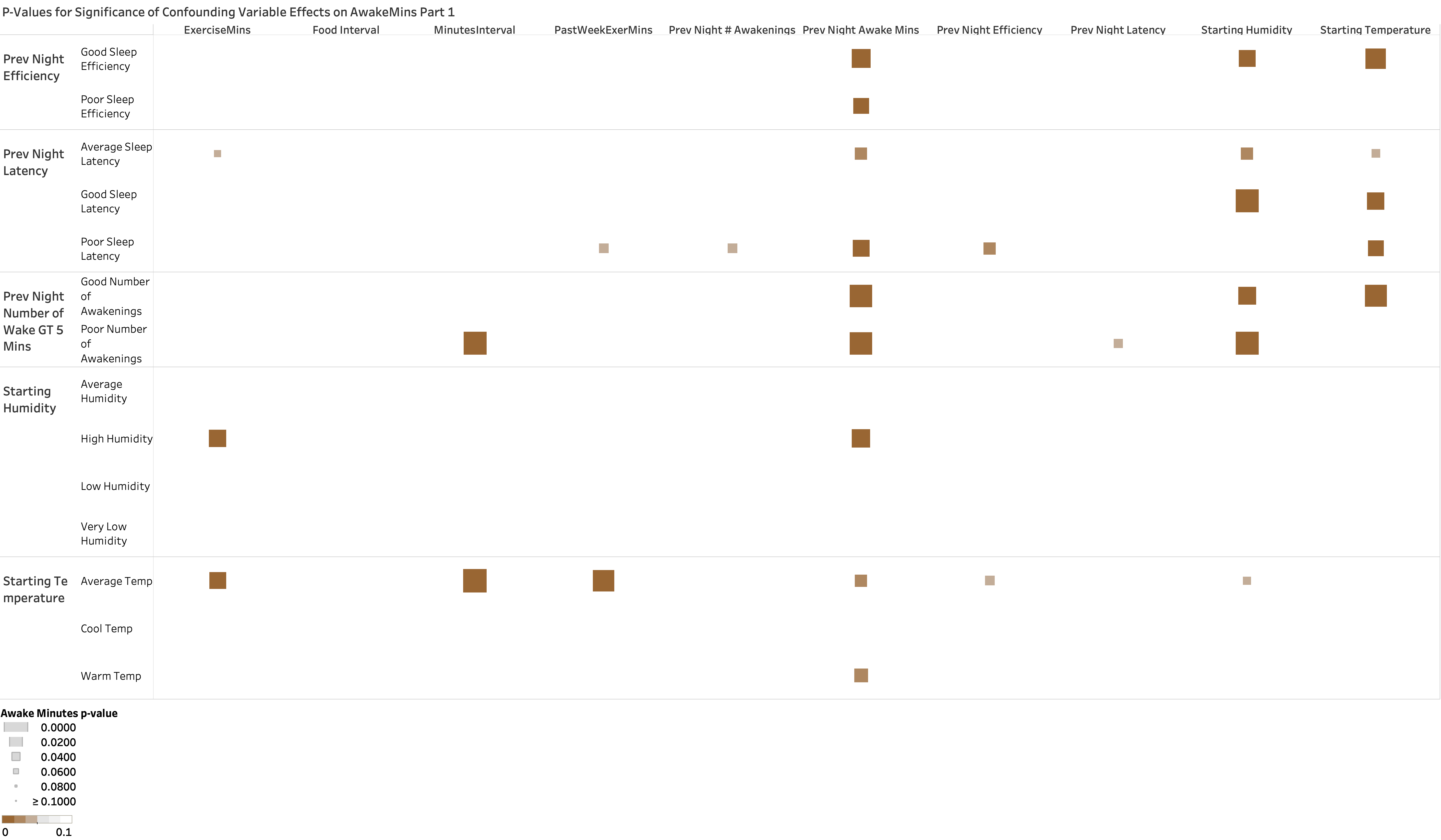}}
\caption{This figure identifies the confounding variables (columns) that affect the relationship between the first half of various input events (rows) and Awake Minutes. The colored squares indicate a significant p-value. The larger the square the more significant the relationship was found.}
\label{fig:AwakePart1}
\end{sidewaysfigure*}

\begin{sidewaysfigure*}
\centerline{\includegraphics[scale=0.175]{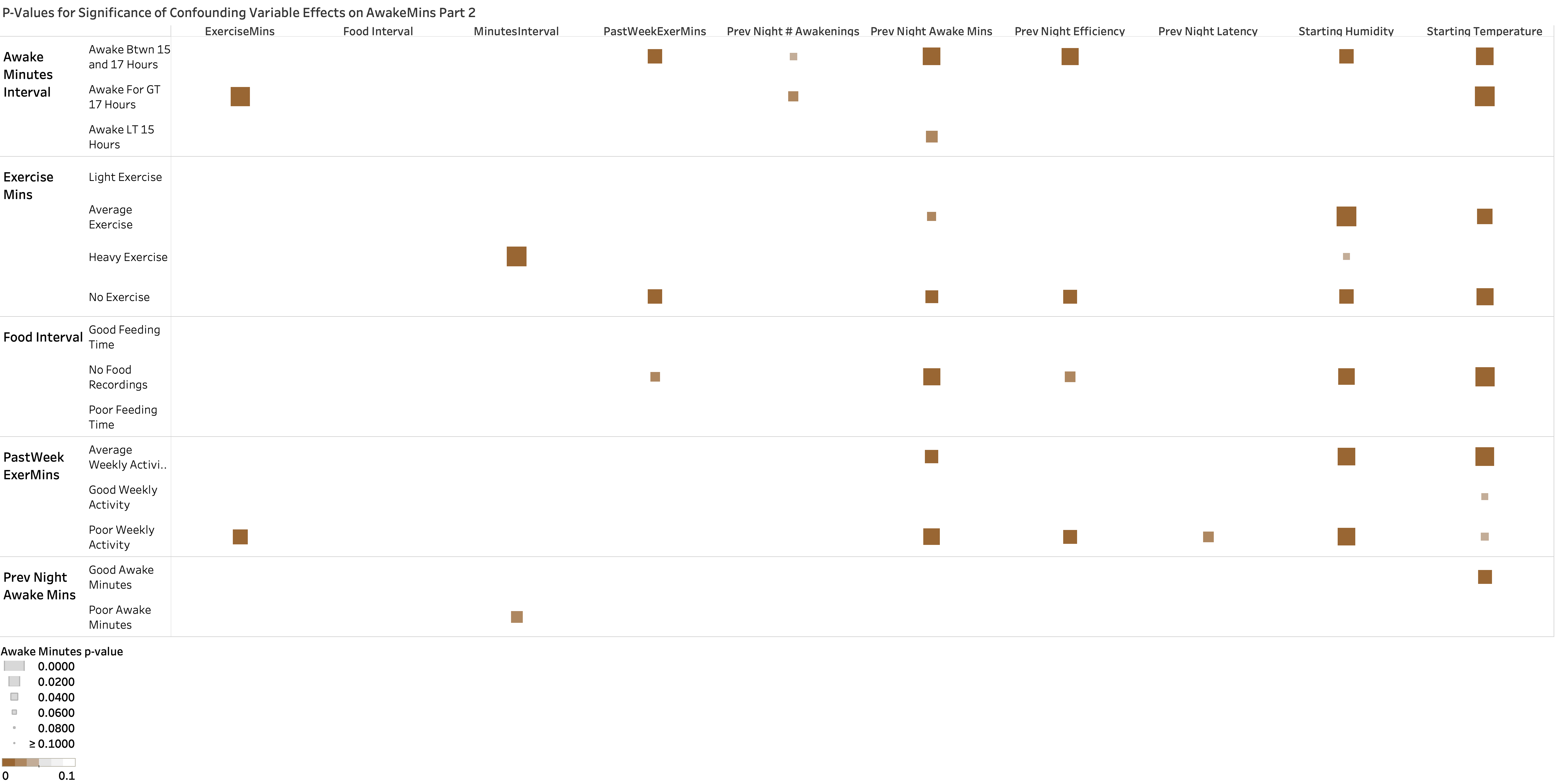}}
\caption{This figure identifies the confounding variables (columns) that affect the relationship between the second half of various input events (rows) and Awake Minutes. The colored squares indicate a significant p-value. The larger the square the more significant the relationship was found.}
\label{fig:AwakePart2}
\end{sidewaysfigure*}

\begin{sidewaysfigure*}
\centerline{\includegraphics[scale=0.175]{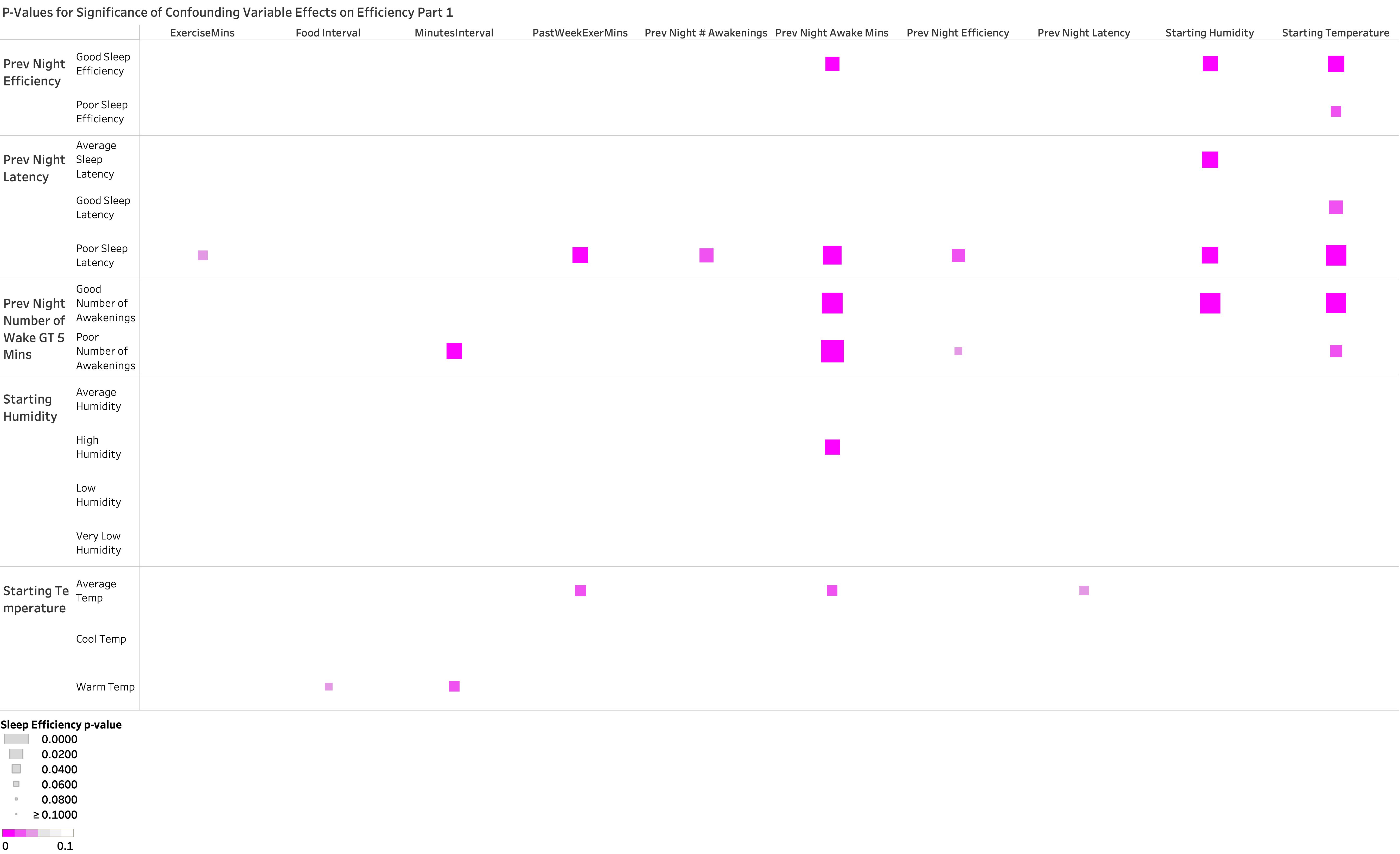}}
\caption{This figure identifies the confounding variables (columns) that affect the relationship between the first half of various input events (rows) and Sleep Efficiency. The colored squares indicate a significant p-value. The larger the square the more significant the relationship was found.}
\label{fig:EffPart1}
\end{sidewaysfigure*}

\begin{sidewaysfigure*}
\centerline{\includegraphics[scale=0.175]{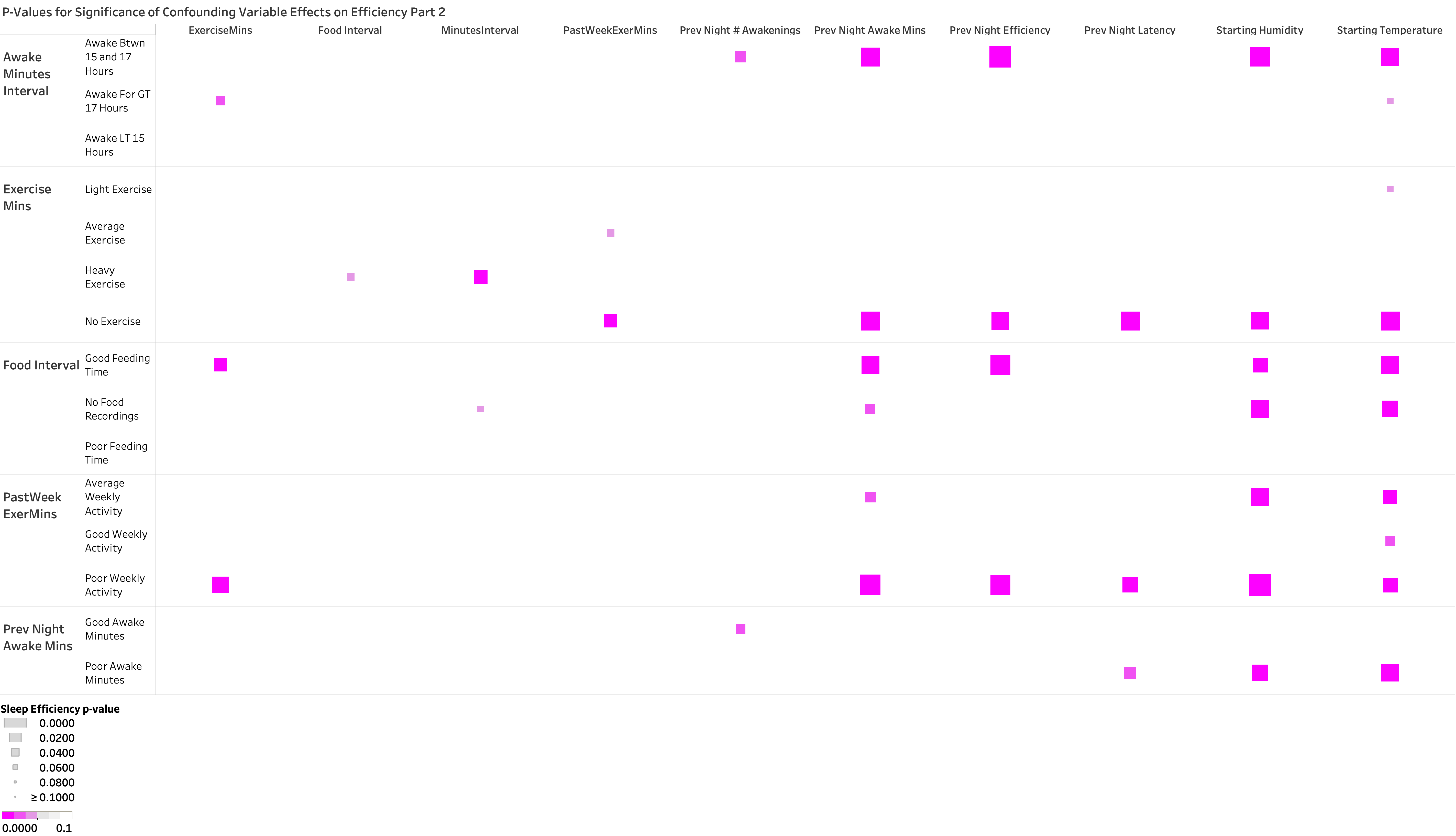}}
\caption{This figure identifies the confounding variables (columns) that affect the relationship between the second half of various input events (rows) and Sleep Efficiency. The colored squares indicate a significant p-value. The larger the square the more significant the relationship was found.}
\label{fig:EffPart2}
\end{sidewaysfigure*}

\end{document}